\documentclass[12pt,reqno]{article}
\usepackage{amsmath,hyperref,amssymb}
\usepackage{subfigure,graphicx,url}

\allowdisplaybreaks \numberwithin{equation}{section}

\DeclareSymbolFont{AMSa}{U}{msa}{m}{n}
\DeclareSymbolFont{AMSb}{U}{msb}{m}{n}
\DeclareMathSymbol{\fieldR}{\mathalpha}{AMSb}{"52}

\begin{document} 
\begin{flushright} \small
 ITP--UU--09/35 \\ SPIN--09/32
\end{flushright}
\bigskip
\begin{center}
 {\large\bfseries Maximally supersymmetric solutions of D=4 N=2 gauged supergravity}\\[5mm]
Kiril Hristov$^{*,\dag}$, Hugo Looyestijn$^*$, Stefan Vandoren$^*$ \\[3mm]
 {\small\slshape
 * Institute for Theoretical Physics \emph{and} Spinoza Institute \\
 Utrecht University, 3508 TD Utrecht, The Netherlands \\
\medskip
 \dag Faculty of Physics, Sofia University, Sofia 1164, Bulgaria\\
\medskip
 {\upshape\ttfamily K.P.Hristov, H.T.Looijestijn, S.J.G.Vandoren@uu.nl}\\[3mm]}
\end{center}
\vspace{5mm} \hrule\bigskip \centerline{\bfseries Abstract}
\medskip

We determine and analyze maximally supersymmetric configurations
in four-dimensional gauged $N=2$ supergravity, preserving eight
supercharges. These models include arbitrary electric gaugings in
the vector- and hypermultiplet sectors. We present several
examples of such solutions and connect some of them to vacuum
solutions of flux compactifications in string theory.

\bigskip
\hrule\bigskip
\section{Introduction}

It is of general interest to study four-dimensional supersymmetric string vacua and
their low-energy effective supergravity descriptions. Firstly, in the context of flux compactifications and
gauged supergravities, one is motivated by the problem of moduli
stabilization and the properties of string vacua in which these
moduli are stabilized. For some reviews on the topic, see
\cite{Grana:2005jc,Douglas:2006es,Blumenhagen:2006ci}. Often, one
focuses on supersymmetric vacua since there is better control over
the dynamics of the theory, though for more realistic situations,
e.g. in accelerating cosmologies, the vacuum must break all
supersymmetry. Secondly, we are motivated to look for new versions
of the $AdS_4$/$CFT_3$ correspondence. The recently proposed
dualities studied in \cite{abjm} are based on $AdS_4$ string vacua
preserving 32 or 24 supersymmetries. Versions of the
$AdS_4$/$CFT_3$ correspondence with less amount of supersymmetry
are not yet well established (for some results on the
correspondence in an $N=2$ setting, see
\cite{gaiotto,klebanov,petrini} and references therein), but are
important for studying aspects of four-dimensional quantum
gravity, and potentially also for certain condensed matter systems
at criticality described by three-dimensional conformal field
theories.

In this paper, we consider four-dimensional $N=2$ gauged
supergravities, and study the configurations that preserve maximal
supersymmetry, i.e. eight supercharges. We only consider electric
gaugings because magnetic gaugings require in addition massive
tensor multiplets which have not been fully constructed yet. In
the ungauged case, $N = 2$ models arise e.g. from Calabi-Yau
compactifications of type II string theories, or $K3\times T^2$
compactifications of the heterotic string. Both models are known
to have a rich dynamical structure with controllable quantum
effects in both vector- and hypermultiplet sectors that are
relatively well understood. Gaugings in $N=2$ supergravity are
well studied and have a long history \cite{de
Wit,Lauwers,preABCDF,ABCDF,D'Auria,deWit:2001bk}. Their analysis in terms of
string compactifications with fluxes started in \cite{Michelson},
and is an ongoing research topic. For a (partial) list of
references, see
\cite{ferrara,kachru,hitchin,aharony,gaunt,cassani,louis}.

In the ungauged case, a complete classification of all the
supersymmetric solutions already exists
\cite{sabra,Ortin,ortinetal}, while there have been also solutions
in the gauged case for (abelian) vector multiplets \cite{klemm}.
We extend this by taking completely general vector- and
hypermultiplet sectors. Since we concentrate only on the maximally
supersymmetric solutions, we use different methods than the ones
in the above references. In fact the space-time conditions we
obtain for our solutions closely resemble other maximally
supersymmetric solutions in different theories such as
\cite{figueroa}.

The plan of the paper is as follows. In section 2, we analyze the
supersymmetry rules and derive the conditions for maximally
supersymmetric vacua. The possible solutions divide in two classes
of space-times, with zero scalar curvature and with negative scalar
curvature, and we explicitly list all the possible outcomes. We
give the lagrangian and the scalar potential for the obtained
vacua in section 3, paying special attention to the
Chern-Simons-like term determined by the $c$-tensor of the
electric gauging. This term generically exists in $N=2$
supergravity and string theory compactifications and we show how
it influences the maximally supersymmetric vacua. In section 4, we
discuss explicit cases from string theory compactifications and
general supergravity considerations that exemplify the use of our
maximal supersymmetry conditions. We have left the definition of
our conventions and notations for the appendices, where we also
present some intermediate and final formulae that are important
for our results.

\section{N=2 supersymmetry rules}

We consider in this section vector multiplets, hypermultiples and
the gravitational multiplet, with arbitrary electric gaugings, and
will mostly follow the notation  of
\cite{ABCDF}, except for some curvature conventions. For completeness,  a list of  conventions is given
in appendix \ref{AppA}.

As is well known, the vector multiplet sector is characterized by
holomorphic sections $X^\Lambda(z)$ and $F_\Lambda(z),
\Lambda=0,1,...,n_V$, and the scalars $z^i; i=1,...,n_V$ parametrize a special
K\"ahler manifold with K\"ahler potential
\begin{equation}\label{K-pot}
{\cal K}(z,\bar z)=-\ln\Big[i({\bar X}^\Lambda(\bar
z)F_\Lambda(z)-X^\Lambda(z) {\bar F}_\Lambda(\bar z))\Big]\ .
\end{equation}
When a prepotential exists, it is given by $2F=X^\Lambda
F_\Lambda$. It should be homogeneous of second degree, and one must have that
$F_\Lambda(X)=\partial F(X)/\partial X^\Lambda$. Our general
analysis does not assume the existence of a prepotential.

The scalars in the hypermultiplet sector parametrize a
quaternion-K\"ahler manifold, whose metric can be expressed in
terms of quaternionic vielbeine. In local coordinates $q^u;
u=1,...,4n_H$, we have
\begin{equation}
h_{uv}(q)=
\mathcal{U}^{A\alpha}_u(q)\,\mathcal{U}^{B\beta}_v(q)\,\mathbb{C}_{\alpha\beta}\,\epsilon_{AB}\
,
\end{equation}
where $\mathbb{C}_{\alpha\beta}, \alpha,\beta=1,...,2n_H$ and
$\epsilon_{AB}, A, B = 1,2$ are the antisymmetric symplectic and
$SU(2)$ metrics, respectively. The value of the Ricci-scalar
curvature of the quaternionic metric is always negative and fixed in terms of Newton's
coupling constant $\kappa$. In units in which $\kappa^2=1$, which
we will use in the remainder of this paper, we have
\begin{equation}
R(h)=-8n_H(n_H+2)\ .
\end{equation}

The analysis of maximally supersymmetric configurations does not
rely on the form of the action, only on the supersymmetry
variations and the equations of motion. Nevertheless, it is
relevant to know what is the value of the scalar potential
evaluated at such a configuration. We therefore turn to the
properties of the Lagrangian in the next section.

It can be seen by inspection that the maximally supersymmetric
configurations\footnote{In this paper we use interchangeably the
terms maximally supersymmetric configurations and BPS
configurations, meaning the field values that are invariant under all eight
supercharges in the theory.} are purely bosonic, and the fermions
need to be zero. This follows from the supersymmetry variations of
the bosonic fields, which can be read off from \cite{ABCDF}.
Therefore, we can restrict ourselves to the supersymmetry
variations of the fermions only.

\subsection{Gauginos}

The number of vector multiplets is denoted by $n_V$, and in $N=2$
special geometry, it is convenient to introduce indices $\Lambda =
0,1,...,n_V$ and $i=1,...,n_V$. The two fermions with positive
chirality in a vector multiplet are denoted by $\lambda^{iA}$,
with $A=1,2$. Complex conjugation changes the chirality and lowers the $SU(2)_R$ indices $A,B,...$\,.
See appendix \ref{AppA} for more on our notations and conventions. Under gauged
supersymmetry, with coupling constant $g$, the gauginos transform
into
\begin{equation}\label{susygluino}
\delta_\varepsilon\lambda^{iA}=i\nabla_\mu z^i
\gamma^\mu\varepsilon^A + G_{\mu\nu}^{-i}
\gamma^{\mu\nu}\epsilon^{AB}\varepsilon_B+gW^{iAB}\varepsilon_B\ ,
\end{equation}
up to terms that are higher order in the fermions and which vanish
for purely bosonic configurations. The supersymmetry parameters are
denoted by $\varepsilon^A$. They have negative chirality and under
complex conjugation $\varepsilon_A\equiv(\varepsilon ^A)^*$,
chirality is flipped since in our conventions $\gamma_5$ is
hermitian but purely imaginary. We explain more on the quantities
in \eqref{susygluino} as we go along.

A maximally supersymmetric configuration preserves the full eight
supercharges, hence the variation of the fermions should vanish
for all choices of the supersymmetry parameters. Since at each
point in spacetime they are linearly independent, the first term
on the right hand side of \eqref{susygluino} must vanish
separately from the others,
\begin{equation}\label{z=covconst}
\nabla_\mu z^i\equiv \partial_\mu z^i + g A_\mu^\Lambda
k^i_\Lambda =0\ .
\end{equation}
It implies the integrability condition\footnote{We will assume in
the remainder of the paper that the gauge coupling constant $g\neq
0$. The case of $g=0$ is treated in the literature in e.g.
\cite{ortinetal}.}
\begin{equation}\label{int-cond-vectors}
F_{\mu\nu}^\Lambda\, k_\Lambda^i =0\ ,
\end{equation}
and complex conjugate. Here, $F^{\Lambda}_{\mu\nu}$ is the full
non-abelian field strength, given by
\begin{align}\label{field-str-def}
  F^\Lambda_{\mu\nu} = \frac 12 (\partial_\mu A_\nu - \partial_\nu
  A_\mu) + \frac 12 f_{\Sigma\Gamma}{}^\Lambda A^\Sigma_\mu
  A^\Gamma_\nu\ .
\end{align}

The $z^i$ are the complex scalars of the vector multiplets, and
$A_\mu^\Lambda$ are the gauge fields (including the graviphoton).
These scalars parametrize a special K\"ahler manifold which may
have a group of isometries. To commute with supersymmetry, these
isometries need to be holomorphic, and we denote the Killing
vector fields by $k_\Lambda(z)$. Under the isometry, the
coordinates change according to
\begin{equation}
\delta_G z^i=-g\alpha^\Lambda k^i_\Lambda(z)\ .
\end{equation}
To close the gauge algebra on the scalars, the Killing vector
fields must span a Lie-algebra with commutation relations
\begin{equation}\label{Lie-alg}
[k_\Lambda,k_\Sigma]=f_{\Lambda\Sigma}{}^{\Pi} k_{\Pi}\ ,
\end{equation}
and structure constants $f_{\Lambda\Sigma}{}^{\Pi}$ of some
Lie-group $G$ that one wishes to gauge. Not all holomorphic
isometries can be gauged within $N=2$ supergravity. The induced
change on the sections needs to be consistent with the symplectic
structure of the theory, and this requires the holomorphic
sections to transform as
\begin{equation}\label{Gtransf-sections}
\delta_G \begin{pmatrix} X^\Lambda \\ F_\Lambda
\end{pmatrix}=-g\alpha^\Sigma\Big[T_\Sigma \cdot\begin{pmatrix}
X^\Lambda \\ F_\Lambda \end{pmatrix}+r_\Sigma(z)\begin{pmatrix}
X^\Lambda \\ F_\Lambda \end{pmatrix}\Big]\ .
\end{equation}
The second term induces a K\"ahler transformation on the K\"ahler
potential
\begin{equation}\label{Gtransf-K}
\delta_G {\cal K}(z,\bar z)=g\alpha^\Lambda(r_\Lambda(z)+{\bar
r}_\Lambda(\bar z))\ ,
\end{equation}
for some holomorphic functions $r_\Lambda(z)$. The first term in
\eqref{Gtransf-sections} contains a constant matrix $T_\Sigma$
that acts on the sections as infinitesimal symplectic
transformations.  For electric gaugings, which we consider in this
section, we mean, by definition, that the representation is of the
form
\begin{equation}\label{el-gauge}
T_\Lambda =\begin{pmatrix} -f_\Lambda & 0 \\ c_\Lambda &
f_\Lambda^t \end{pmatrix}\ ,
\end{equation}
where $f_\Lambda$ denotes the matrix
$(f_\Lambda)_{\Sigma}{}^\Pi=f_{\Lambda\Sigma}{}^\Pi$ and
$f^t_\Lambda$ is the transposed. The tensor
$c_{\Lambda,\Sigma\Pi}\equiv (c_\Lambda)_{\Sigma\Pi}$ is required
to be symmetric for $T_\Lambda$ to be a symplectic generator.
Moreover, there are some additional constraints on the $c_\Lambda$
in order for the $T_\Lambda$ to be symplectically embedded within
the same Lie-algebra as in
\eqref{Lie-alg}. One can easily derive them, for explicit formulae
see \cite{Lauwers}, or \eqref{struct const condition}. Finally, closure of the gauge transformations on the
K\"ahler potential requires that
\begin{equation}\label{equiv2}
k^i_\Lambda \partial_ir_\Sigma - k^i_\Sigma \partial_ir_\Lambda =
f_{\Lambda\Sigma}{}^\Pi r_\Pi\ .
\end{equation}
We summarize some other important identities on vector multiplet
gauging in appendix \ref{AppB}.

Magnetic gaugings allow also non-zero entries in the upper--right
corner of $T_\Lambda$, but we will not consider them here. The
gauged action, in particular the scalar potential, that we
consider below is not invariant under magnetic gauge
transformation. To restore this invariance, one needs to introduce
massive tensor multiplets, but the most general lagrangian with
both electric and magnetic gauging is not fully understood yet
(for some partial results see
\cite{scal-tens,tens2,bernard,it-tens}).

Given a choice for the gauge group \eqref{el-gauge}, one can
reverse the order of logic and determine the form of the Killing
vectors, and therefore the gauge transformations of the scalar
fields $z^i$. This analysis was done in \cite{vProeyen}, and the
result is written in the appendix, see \eqref{Kill-vect}.

We now return to the BPS conditions.
The second and third term in the supersymmetry variation of the
gauginos, equation \eqref{susygluino}, need also to vanish
separately, since they multiply independent spinors of the same
chirality. For the second term, this leads to
\begin{equation}\label{defG}
G_{\mu\nu}^{i\,-}\equiv -g^{i\bar \jmath}{\bar f}_{\bar
\jmath}^\Lambda ({\rm Im} {\cal
N}_{\Lambda\Sigma})F_{\mu\nu}^{\Sigma\,-}=0\ ,
\end{equation}
where $g^{i\bar \jmath}$ is the inverse K\"ahler metric with
K\"ahler potential ${\cal K}$ from \eqref{K-pot}, and
\begin{equation}\label{period-matrix}
{\overline {\cal N}}_{\Lambda \Sigma}\equiv \begin{pmatrix} D_iF_\Lambda \\
{\bar F}_\Lambda\end{pmatrix} \cdot {\begin{pmatrix} D_i X^\Sigma \\
{\bar X}^\Sigma\end{pmatrix}}^{-1} \ ,\qquad f^\Lambda_i\equiv{\rm
e}^{{\cal K}/2}D_i X^\Lambda\ ,
\end{equation}
with $D_iX^\Lambda = (\partial_i+{\cal K}_i)X^\Lambda$ and
similarly $D_iF_\Lambda = (\partial_i+{\cal K}_i)F_\Lambda$. The
anti-selfdual part of any real two-form $T_{\mu\nu}$ is denoted by
$T_{\mu\nu}^-$, and complex conjugation gives the selfdual part,
see the appendix of \cite{ABCDF}.

Finally, setting the third term in the supersymmetry variation to
zero leads to
\begin{equation}\label{defW}
W^{iAB}\equiv k^i_\Lambda {\bar L}^\Lambda \epsilon^{AB} + i
g^{i\bar \jmath}{\bar f}^\Lambda_{\bar
\jmath}P^x_\Lambda\sigma_x^{AB}=0\ ,
\end{equation}
where $L^\Lambda = {\rm e}^{{\cal K}/2}X^\Lambda$ (in analogy,
$M_{\Lambda} \equiv {\rm e}^{{\cal K}/2} F_{\Lambda}$) and $P^x_\Lambda$
are the triplet of moment maps associated with the Killing vector
fields ${\tilde k}_\Lambda$ on the quaternionic
geometry\footnote{For the explicit relation between moment maps
and Killing vectors in the quaternionic case, as well as other
useful identities in the hypermultiplet sector, see the standard
references.}. These Killing vectors are used to determine the
gauge transformations of the hypermultiplet scalars under the
gauge group. The only requirement is that the Killing equation is
satisfied, i.e. they are isometries on the quaternion-K\"ahler
manifold, and they satisfy the same Lie-bracket as in
\eqref{Lie-alg}. Of course, a given quaternion-K\"ahler manifold
can allow inequivalent choices of Killing vectors with the same
Lie-algebra. These choices lead to different models with different
physics. One obvious choice is to set all the Killing vectors to
zero, and so all hypermultiplet scalars remain neutral under the
gauge group. The gauging then remains solely active on the vector
multiplet scalars.

Close inspection of \eqref{defW} shows that both terms are
linearly independent in $SU(2)_R$ space, hence they must vanish
separately,
\begin{equation}\label{Pf}
k^i_\Lambda {\bar L}^\Lambda =0 \ ,\qquad P^x_\Lambda
f^\Lambda_i=0\ ,
\end{equation}
and their complex conjugates.

\subsection{Hyperinos}\label{hyperinos}

The fields in the hypermultiplet sector comprise $4n_H$ scalars
$q^u$, and $2n_H$ positive chirality fermions $\zeta_\alpha$ and
their complex conjugates
$(\zeta_\alpha)^*=\mathbb{C}_{\alpha\beta}\zeta^\beta$ with
negative chirality. Under $N=2$ local supersymmetry, these
hyperinos transform as
\begin{equation}\label{susy-hyperino}
\delta_\varepsilon \zeta_\alpha = i\,
\mathcal{U}^{B\beta}_u\nabla_\mu q^u \gamma^\mu \varepsilon^A
\epsilon_{AB}\mathbb{C}_{\alpha\beta} + g N_\alpha^A\varepsilon_A
\ ,
\end{equation}
again, up to terms that are of higher order in the fermions. The
hyperino mass matrix $N^A_\alpha$ is defined by
\begin{equation}
N_\alpha^A\equiv 2\,\mathcal{U}^A_{\alpha\,u}{\tilde k}^u_\Lambda
{\bar L}^\Lambda\ ,
\end{equation}
with $L^\Lambda$ as given just below \eqref{defW}.

Similarly as for the gauginos, $N=2$ supersymmetric configurations
 require the two terms in \eqref{susy-hyperino} to vanish separately.
Since the quaternionic vielbeine are invertible and nowhere
vanishing, the scalars need to be covariantly constant,
\begin{equation}
\nabla_\mu q^u\equiv \partial_\mu q^u + g A_\mu^\Lambda {\tilde
k}_\Lambda^u =0\ ,
\end{equation}
implying the integrability conditions
\begin{equation}\label{int-cond-hypers}
F_{\mu\nu}^\Lambda\, {\tilde k}_\Lambda^u=0\ .
\end{equation}
Furthermore, there is a second condition from
\eqref{susy-hyperino} coming from the vanishing of the hyperino
mass matrix $N_\alpha^A$. This leads to
\begin{equation}\label{kHL=0}
{\tilde k}^u_\Lambda L^\Lambda =0\ ,
\end{equation}
and complex conjugate.

In the absence of hypermultiplets, i.e. when $n_H=0$,
the $N=2$ conditions from the variations of the hyperinos
disappear. However, the second condition in \eqref{Pf} remains, with
the moment maps replaced by FI parameters\footnote{In
the absence of any hypermultiplets the quantities $P^x_{\Lambda}$
need not vanish. Instead, they can be constants, which can be non-vanishing for
gauge groups $SU(2)$ or $U(1)$. These constants are sometimes
referred to as Fayet-Illiopoulos (FI) terms. See e.g. \cite{Bergshoeff:2004kh} for a discussion.}. So our formalism
automatically includes the case $n_H=0$.

\subsection{Gravitinos}

The fermions in the gravitational sector are two gravitinos of
opposite chirality $\psi_{\mu A}$ and its complex conjugate
$\psi_\mu^A=(\psi_{\mu A})^*$. In gauged supergravity, their
supersymmetry transformation rules are (up to irrelevant higher
order terms in the fermions)
\begin{equation}\label{susy-gravi}
\delta_\varepsilon \psi_{\mu A}=\nabla_\mu\varepsilon_A +
T^-_{\mu\nu}\gamma^\nu
\epsilon_{AB}\varepsilon^B+igS_{AB}\gamma_\mu\varepsilon^B\ .
\end{equation}
Here, $\nabla_\mu\varepsilon_A$ is the gauged supercovariant
derivative (specified below), and
\begin{equation}
T^-_{\mu\nu}\equiv 2iF^{\Lambda\,-}_{\mu\nu}({\rm Im}{\cal
N}_{\Lambda\Sigma}) L^\Sigma \ ,\qquad
S_{AB}\equiv\frac{i}{2}(\sigma_x)_{AB}P^x_\Lambda L^\Lambda\ .
\end{equation}
The matrices $T_{\mu\nu}$ and $S_{AB}$ are called the graviphoton
field strength and the gravitino mass-matrix respectively. Notice
again that for $n_H=0$, in fact even also in the absence of vector
multiplets when $n_V=0$, the gravitino mass-matrix can be
non-vanishing and constant. In the Lagrangian, which we discuss in
the next section, this leads to a (negative) cosmological constant
term. The anti-selfdual part of the graviphoton field strength
$T_{\mu\nu}$ satisfies the identity
\begin{equation} \label{graviphoton-identity}
F_{\mu\nu}^{\Lambda\,-}=i{\bar L}^\Lambda
T^-_{\mu\nu}+2f_i^\Lambda G^{i\,-}_{\mu\nu}\ ,
\end{equation}
with $G_{\mu\nu}^{i\,-}$ defined in \eqref{defG}. From the
vanishing of the gaugino variation, we have that
$G_{\mu\nu}^{i\,-}=0$, and hence a maximally supersymmetric
configuration must satisfy $F_{\mu\nu}^{\Lambda\,-}=i{\bar
L}^\Lambda T^-_{\mu\nu}$, or
\begin{equation}\label{F-LT}
  F_{\mu \nu}^\Lambda = i \overline L^\Lambda T^-_{\mu \nu} - i
  L^\Lambda T^+_{\mu \nu}\ .
\end{equation}
Using this, we then see that equation~\eqref{kHL=0} implies
the integrability conditions~\eqref{int-cond-hypers} in the
hypermultiplet sector. For the integrability equations in the
vector multiplet sector, the situation is more subtle, as the
Killing vectors are complex and holomorphic. Now, the BPS
condition \eqref{Pf} only implies that
\begin{align}\label{int-relation}
  k^i_\Lambda F^\Lambda_{\mu \nu} = -i k^i_\Lambda L^\Lambda T^+_{\mu
    \nu}\ .
\end{align}
As a consequence, the integrability condition
\eqref{int-cond-vectors} is only guaranteed when $k^i_\Lambda
L^\Lambda = 0$ (or, when $T_{\mu\nu}=0$, but then all the field
strengths are zero). So, for $T_{\mu\nu}\neq 0$, a necessary
condition for a maximally supersymmetric configuration is that
$k^i_\Lambda L^\Lambda =0$. Furthermore, in appendix \ref{AppB} we
prove that
\begin{equation}\label{PL=0}
k^i_\Lambda L^\Lambda = 0  \quad \Leftrightarrow \quad P_\Lambda
L^\Lambda =0\,
\end{equation}
where $P_\Lambda$ is the special K\"ahler moment map, defined in
\eqref{mom-maps}.

\textbf{Note added:} In fact, we show in appendix B that $P_\Lambda L^\Lambda = 0$ is an identity of the theory,
and hence the integrability condition is always satisfied.

In terms of \eqref{mom-maps2}, one sees that this condition
involves both the structure constants and the matrix $c_\Lambda$.
Hence the integrability condition is satisfied for those
configurations satisfying $P_\Lambda L^\Lambda =0$. The integrability condition
might only locally be sufficient, but this fine for our purposes. One might however check in
addition whether the covariant constancy of the vector multiplet scalars imposes further (global) restrictions.

To solve the constraints from the gravitino variation, we must
first specify the gauged supercovariant derivative on the
supersymmetry parameter. It can be written as
\begin{equation}\label{scov-eps}
\nabla_\mu\varepsilon_A=(\partial_\mu -\frac{1}{4}{\cal
\omega}_\mu^{ab}\gamma_{ab})\varepsilon_A+\frac{i}{2} A_\mu
\varepsilon_A+\omega_{\mu\,A}{}^B\varepsilon_B\ .
\end{equation}
The conventions for the spin connection, appearing between the
brackets, are specified in the appendix. Furthermore, there appear
two other  connections associated to the special K\"ahler and
quaternion-K\"ahler manifolds. We need to compute their curvatures
since they enter the integrability conditions that
follow from the Killing spinor equations. The first one is called
the gauged $U(1)$ K\"ahler-connection, defined by
\cite{ABCDF,vProeyen}
\begin{equation}\label{gaugedU1}
A_\mu\equiv -\frac{i}{2}\Big(\partial_i{\cal K}\nabla_\mu
z^i-\partial_{\bar\iota}{\cal K}\nabla_\mu{\bar z}^{\bar
\iota}\Big)-\frac{i}{2}gA_\mu^\Lambda (r_\Lambda -{\bar
r}_\Lambda)\ .
\end{equation}
Under a gauge transformation, one finds that
\begin{equation}
\delta_G A_\mu =
\frac{i}{2}g\,\partial_\mu\Big[\alpha^\Lambda(r_\Lambda -{\bar
r}_\Lambda)\Big]\ .
\end{equation}
The curvature of this connection can be computed to be
\begin{equation}
F_{\mu\nu} =ig_{i\bar\jmath}\nabla_{[\mu}z^i\nabla_{\nu ]}{\bar
z}^{\bar\jmath}-gF_{\mu\nu}^\Lambda P_\Lambda\ ,
\end{equation}
where $P_\Lambda$ is the moment map, defined in \eqref{mom-maps},
and we have used the equivariance condition \eqref{equivar}. For
maximally supersymmetric configurations, the scalars are
covariantly constant and hence the curvature of the K\"ahler
connections satisfies $F_{\mu\nu}=-gF_{\mu\nu}^\Lambda P_\Lambda$.

The second connection appearing in the gravitino supersymmetry
variation is the gauged $Sp(1)$ connection of the
quaternion-K\"ahler manifold. It reads
\begin{equation}\label{gauged-Sp1}
 \omega_{\mu \,A}{}^B\equiv\partial_\mu q^u \omega_{u\,A}{}^B+gA_\mu^\Lambda P_{\Lambda\,A}{}^B\ ,
 \end{equation}
 where  $\omega_{u\,A}{}^B$ is the (ungauged) $Sp(1)$ connection of the quaternion-K\"ahler manifold, whose curvatures are related to the three quaternionic two-forms. The effect of the gauging is to add the second term on the right hand side of \eqref{gauged-Sp1}, proportional to the triplet of moment maps of the quaternionic isometries,
 with $P_{\Lambda\,A}{}^B= \frac i 2 P^x_\Lambda (\sigma^x)_A{}^B$. The curvature of \eqref{gauged-Sp1} can then be computed to be
 \begin{align}
 \Omega_{\mu\nu\,A}{}^B =2\Omega_{uv\,A}{}^B\nabla_{[\mu}q^u\nabla_{\nu]}q^v+gF_{\mu\nu}^\Lambda P_{\Lambda\,A}{}^B\ ,
 \end{align}
where $\Omega_{uv\,A}{}^B$ is the quaternionic curvature.
 For
fully BPS solutions, we have
$\Omega_{\mu\nu\,A}{}^B=gF_{\mu\nu}^\Lambda P_{\Lambda\,A}{}^B$.

We can now investigate the integrability  conditions that follow
from the vanishing of the gravitino transformation rules
\eqref{susy-gravi}. From the definition of the supercovariant
derivative \eqref{scov-eps}, we find\footnote{Strictly speaking,
we get the supercovariant curvatures appearing in
\eqref{comm-eps}, which also contain fermion bilinears. Since the
fermions are zero on maximally supersymmetric configurations, only
the bosonic part of the curvatures remains.}
 \begin{equation}\label{comm-eps}
 [\nabla_\mu,\nabla_\nu]\varepsilon_A=-\frac{1}{4}R_{\mu\nu}{}^{ab}\gamma_{ab}\,\varepsilon_A-
 ig F_{\mu\nu}^\Lambda P_\Lambda\varepsilon_A + 2gF_{\mu\nu}^\Lambda P_{\Lambda\,A}{}^B\varepsilon_B\ ,
 \end{equation}
 where we have used the covariant constancy of the scalars. We remind that $P_\Lambda$ are the moment maps on the special K\"ahler geometry, whereas $P_{\Lambda\,A}{}^B$ are the quaternion-K\"ahler moment maps. Alternatively, we can compute the commutator from
 the vanishing of the gravitino variations spelled out in \eqref{susy-gravi}. By equating this to the result of \eqref{comm-eps}, we get a set of constraints.
 Details of the calculation are given in appendix \ref{AppC}, and the results can be summarized as follows. First of all, we find the covariant constancy of the graviphoton field
 strength\footnote{Recall that $T^+$ and $T^-$ are related by complex conjugation, and hence the vanishing of $DT^+$ implies $DT=0$.}
 \begin{equation}\label{cov-cons-graviphoton}
D_\rho T_{\mu\nu}^+=0\ .
\end{equation}
Secondly, we get that the quaternionic moment maps must satisfy
\begin{equation}
\epsilon^{xyz}P^y \overline {P^z}=0\ ,\qquad P^x\equiv L^\Lambda
P_\Lambda^x\ .
\end{equation}
Moreover, there are cross terms between the graviphoton and the
moment maps, which enforce the conditions
\begin{equation}
T^+_{\mu\nu}\,P^x=0\ .
\end{equation}
This equation separates the classification of BPS configurations
in two sectors, those with a solution of $P^x=0$ at a particular
point (or locus) in field space, and those with non-vanishing
$P^x$ (for at least one index $x$) but $T_{\mu\nu}=0$. We will see
later on that this distinction corresponds to zero or non-zero
(and negative) cosmological constant in the spacetime.

Another requirement that follows from the gravitino integrability
conditions is
\begin{equation}\label{int-field-moment}
F_{\mu\nu}^\Lambda P_\Lambda =0\ ,
\end{equation}
where $P_\Lambda$ is defined in \eqref{mom-maps}, and is real. Using
\eqref{F-LT}, this is equivalent to the condition
\begin{equation}
{\bar L}^\Lambda P_\Lambda T^-_{\mu\nu}=L^\Lambda P_\Lambda T^+_{\mu\nu}\ .
\end{equation}
Since anti-selfdual and selfdual tensors are linearly independent,
it means that
 $P_\Lambda L^\Lambda =0$ and complex conjugate (again, for $T_{\mu\nu}\neq 0$). This
requirement is already imposed by the integrability conditions on the
vector multiplet scalars, see \eqref{PL=0}, so
\eqref{int-field-moment} does not lead to any new constraint.

Finally, there is the condition on the spacetime Riemann
curvature. It reads
\begin{equation}\label{spacetime-Riemann}
R_{\mu\nu\rho\sigma}=4T^+_{\mu [\sigma}T^-_{\rho
]\nu}+g^2P^x{\overline
{P^x}}g_{\mu\sigma}g_{\nu\rho}-(\mu\leftrightarrow \nu)\ .
\end{equation}
It can be checked that this leads to a vanishing Weyl tensor, implying
conformal flatness. From the curvature, we can compute the value of the Ricci-scalar to be
\begin{equation}
R=-12g^2 P^x{\overline {P^x}}\ .
\end{equation}
Hence, the classification of fully supersymmetric configurations
separates into negative scalar curvature with $P^x \overline
{P^x}\neq 0$, and zero curvature with $P^x=0$ at the
supersymmetric point. In both of these cases there are important
simplifications.

\subsubsection{Negative scalar curvature}
The case of negative scalar curvature is characterized by
$T_{\mu\nu}=0$ and $P^x \overline {P^x} \neq 0$ at the supersymmetric point. Since the BPS
conditions imply that then both $T_{\mu\nu}$ and
$G^{i-}_{\mu\nu}=0$ (see equation~\eqref{defG}), we find that all
field strengths should be zero: $F^\Lambda_{\mu \nu} = 0$. The
gauge fields then are required to be pure gauge, but can still be
topologically non-trivial. Furthermore, because of the vanishing field strengths,
the integrability conditions on the scalar fields are satisfied,
and a solution for the sections  $X^\Lambda(z)$ is obtained by a
gauge transformation on the constant (in spacetime) sections.
Finally, the Riemann tensor is  given by
\begin{align*}
R_{\mu\nu\rho\sigma}&=g^2P^x{\overline
{P^x}}\left(g_{\mu\sigma}g_{\nu\rho}-g_{\nu\sigma}g_{\mu\rho}\right)\
.
\end{align*}
which shows that the space is maximally symmetric, and therefore locally $AdS_4$. The scalar
curvature is $R=-12g^2 P^x \overline {P^x}$.

\subsubsection{Zero scalar curvature}
The class of zero curvature is characterized by configurations for
which $P^x=0$ at the supersymmetric point.  In this case, we can combine the conditions
$P^x_\Lambda f_i^\Lambda = 0$ and $P^x \equiv P^x_\Lambda
L^\Lambda =0$ into
\begin{align*}
  P^x_\Lambda \begin{pmatrix} \bar L^\Lambda \\
    f_i^\Lambda \end{pmatrix} =0\ .
\end{align*}
The matrix appearing here is the invertible matrix of special geometry (as used in~\eqref{period-matrix}), hence
we conclude that $P^x_\Lambda = 0$. The Riemann tensor is then
\begin{align*}
R_{\mu\nu\rho\sigma}&=4T^+_{\mu [\sigma}T^-_{\rho
  ]\nu}-(\mu\leftrightarrow \nu)\ .
\end{align*}
From the covariant constancy of the graviphoton,
condition~\eqref{cov-cons-graviphoton}, we find $D_\rho R_{\mu \nu
\sigma\tau} = 0$. Spaces with covariantly constant Riemann tensor
are called locally symmetric, and they are classified, see e.g.
\cite{CahenWallach,Ortin,figueroa}. In our case we also have zero
scalar curvature, and then only three spaces are possible:
\begin{enumerate}
\item Minkowski space $M_4$ ($T_{\mu\nu}=0$)
\item $AdS_2 \times S^2$
\item The pp-wave solution
\end{enumerate}
The explicit metrics and field strengths for the latter two cases  ($M_4$ and $AdS_4$ are well-known and have vanishing
field strengths) are listed in appendix \ref{AppD}.

\subsection{Summary}\label{summary}
Let us now summarize the results. There are two different classes:
negative scalar curvature (leading to $AdS_4$) and zero scalar
curvature solutions (leading to $M_4$, $AdS_2 \times S^2$ or the pp--wave).

The result of our analysis is that all the conditions on the
spacetime dependent part are explicitly solved\footnote{This is apart from
the scalar fields and Killing spinors, which are spacetime
dependent. The integrability conditions that we have imposed
guarantee locally the existence of a solution, although we did not
explicitly construct it. Its construction cannot be done in closed
form in full generality, but can be worked out in any given
example \cite{Ortin}.}, and the remaining conditions are
purely algebraic, and depend only on the geometry of the special
K\"ahler and quaternionic  manifolds. The solutions to these
algebraic equations define the configuration space of maximally
supersymmetric configurations. There are two separate cases:

\subsubsection{ Negative scalar curvature ($AdS_4$)}\label{2.4.1}
This case is characterized by configurations for which $P^x
\overline{P^x} \neq 0$ at the supersymmetric point. The BPS conditions are
\newcommand{\boxedeqn}[1]{%
  \[\fbox{%
      \addtolength{\linewidth}{-2\fboxsep}%
      \addtolength{\linewidth}{-2\fboxrule}%
      \begin{minipage}{\linewidth}%
      \begin{align*}#1\end{align*}%
      \end{minipage}%
    }\]%
} \boxedeqn{ k^i_\Lambda \overline L^\Lambda & =0 &
\tilde k^u_\Lambda L^\Lambda &=0 \\
P^x_\Lambda f_i^\Lambda &=0 &  \epsilon^{xyz} P^y \overline {P^z} &= 0\ ,\\
} \\
which should be satisfied at a point (or a locus) in field
space. The field strengths are zero, $F^\Lambda_{\mu \nu}=0$, and
the space--time is $AdS_4$ with scalar curvature $R = -12 g^2 P^x
\overline{P^x}$.

\subsubsection{Zero scalar curvature ($M_4$, $AdS_2 \times
  S^2$ or pp--wave)}\label{2.4.2}
In this case, the BPS conditions are \boxedeqn{
k^i_\Lambda \overline L^\Lambda &=0 & \tilde k^u_\Lambda L^\Lambda &=0 \\
P_\Lambda L^\Lambda    &= 0 &
P^x_\Lambda &=0\ .\\
} We remind that, when $T_{\mu\nu}=0$ (Minkowski space), all field
strengths are vanishing ($F_{\mu\nu}^\Lambda=0$), and the
condition $P_\Lambda L^\Lambda=0$ need not be satisfied. For
non-vanishing $T_{\mu\nu}$, the field strengths are given by
\eqref{F-LT}, and using formula~\eqref{mom-maps2} the condition
$P_\Lambda L^\Lambda=0$ is equivalent to
\begin{equation}\label{constr on pp-wave and r-b}
 L^\Lambda {\bar L}^\Pi\,  f_{\Lambda \Pi}{}^\Sigma M_\Sigma  +
  L^\Lambda L^\Pi\,c_{\Lambda,\Pi \Sigma}\,  {\bar L}^\Sigma=0\ ,
\end{equation}
where we remind that $M_\Lambda \equiv {\rm e}^{{\cal K}/2}F_\Lambda$.
Hence the existence of maximal BPS configurations also depends on
the $c_\Lambda$-matrix characterizing the Chern-Simons-like terms.

\textbf{Note added:} In fact, we show in appendix B that $P_\Lambda L^\Lambda = 0$
is an identity of the theory, and hence can be removed from the list
of BPS conditions.

\section{Lagrangians and scalar potentials}

Since all fermions are equal to zero for $N=2$ supersymmetric
configurations, we can concentrate on the bosonic part of the
Lagrangian, with action $S=\int\,{\rm d}^4x\,{\sqrt g}{\cal L}$.
It can be read off from \cite{Lauwers,ABCDF},
\begin{eqnarray}\label{lagr}
{\cal L}&=&\frac{1}{2}R(g)+g_{i\bar \jmath}\nabla^\mu z^i
\nabla_\mu
{\bar z}^{\bar \jmath} + h_{uv}\nabla^\mu q^u \nabla_\mu q^v + ({\rm Im}\,{\cal N}_{\Lambda\Sigma})F_{\mu\nu}^{\Lambda}F^{\Sigma\,\mu\nu}\\
&&+\frac{1}{2}({\rm Re}\,{\cal
N}_{\Lambda\Sigma})\epsilon^{\mu\nu\rho\sigma}
F_{\mu\nu}^{\Lambda}F^{\Sigma}_{\rho\sigma}-\frac 13 g\,c_{\Lambda,\Sigma\Pi}\,\epsilon^{\mu\nu\rho\sigma}A_\mu^\Lambda
A_\nu^\Sigma \left( \partial_\rho
A_\sigma^\Pi-\frac{3}{8}f_{\Omega\Gamma}{}^\Pi A_\rho^\Omega
A_\sigma^\Gamma \right) -V(z,\bar z,q)\ ,\nonumber
\end{eqnarray}
with scalar potential
\begin{equation}\label{pot2}
V=g^2\Big[ (g_{i\bar \jmath}k^i_\Lambda k^{\bar \jmath}_\Sigma + 4
h_{uv}k^u_\Lambda k^v_\Sigma){\bar L}^\Lambda L^\Sigma + (g^{i\bar
\jmath}f_i^\Lambda {\bar f}_{\bar \jmath}^\Sigma -3{\bar
L}^\Lambda L^\Sigma)P^x_\Lambda P^x_{\Sigma}\Big]\ .
\end{equation}
The Chern-Simons-like term on the second line of \eqref{lagr} can
be determined from the non gauge--invariance of the period matrix.
From \eqref{period-matrix} one finds
\begin{equation}
\delta_G {\cal N}_{\Lambda\Sigma}=-g\alpha^\Pi \left( f_{\Pi
\Lambda}{}^{\Gamma} {\cal N}_{\Gamma \Sigma} + f_{\Pi
\Sigma}{}^{\Gamma} {\cal N}_{\Gamma \Lambda} + c_{\Pi,\Lambda
\Sigma} \right).
\end{equation}
Since the last term on the right hand side is real, the
topological term proportional to ${\rm Re}\,{\cal
N}_{\Lambda\Sigma}$ in the action is not gauge invariant. This is
compensated by the gauge transformation of the other terms in the
second line, using the various constraints on the (symmetric)
$c_\Lambda$. In the abelian case, the only constraint is that the
totally symmetrized $c$-tensor vanishes, i.e.
\begin{equation}\label{cyclicity}
c_{\Lambda,\Sigma\Pi}+c_{\Pi,\Lambda\Sigma}+c_{\Sigma,\Pi\Lambda}=0\
.
\end{equation}
This implies that for a single vector field, the Chern-Simons-like
term vanishes. The additional constraints for nonabelian gaugings
involve the structure constants \cite{Lauwers}:
\begin{equation}\label{struct const condition}
  f_{\Lambda \Sigma}{}^{\Gamma} c_{\Gamma, \Pi \Omega} + f_{\Omega
  \Sigma}{}^{\Gamma} c_{\Lambda, \Gamma \Pi} + f_{\Pi
  \Sigma}{}^{\Gamma} c_{\Lambda, \Gamma \Omega} + f_{\Lambda
  \Omega}{}^{\Gamma} c_{\Sigma, \Gamma \Pi} + f_{\Lambda
  \Pi}{}^{\Gamma}c_{\Sigma, \Gamma \Omega} = 0\ .
\end{equation}

The scalar potential can be written in terms of the mass-matrices,
\begin{equation}\label{potential}
V=-6S^{AB}S_{AB}+\frac{1}{2}g_{i\bar \jmath}W^{iAB}{\overline
W}^{\bar \jmath}_{AB}+N_\alpha^A N^\alpha_A\ .
\end{equation}
Since the gaugino and hyperino mass-matrices, $W^{iAB}$ and
$N^A_\alpha$ respectively, vanish on $N=2$ supersymmetric
configurations, one sees that the scalar potential is
semi-negative definite, and determined by the gravitino
mass-matrix $S_{AB}$. Even in the absence of vector and
hypermultiplets, the gravitino mass-matrix can be non-vanishing,
leading to a negative cosmological constant in the Lagrangian.
Using \eqref{pot2}, we find for $N=2$ preserving configurations
\begin{equation}
V=-3g^2{\bar L}^\Lambda L^\Sigma P^x_\Lambda P^x_{\Sigma}\ .
\end{equation}
In the absence of hypermultiplets, $N=2$ preserving $AdS_4$ vacua
can therefore only be generated by non-trivial Fayet-Illiopoulos terms.

It can be verified that maximally supersymmetric configurations also solve the equations of motion.
To show this, one varies
the lagrangian~\eqref{lagr} and uses the
identities~\eqref{cyclicity}, \eqref{struct const condition} and
the formulas in section~\ref{summary}. After a somewhat tedious but
straightforward computation one sees that all equations of motion
are indeed satisfied by the maximally supersymmetric
configurations.

\section{Examples}
In this section we list some (string theory motivated) examples of
$N=2, D=4$ theories, leading to $N=2$ supersymmetric configurations.
We will first mention briefly
some already known and relatively well-understood $N=2$ vacua from
string theory and then concentrate on our two main examples in
subsections \ref{su(3) compactification} and \ref{SE_7} that
exhibit best the different features discussed above. In the last
subsection we include some supergravity models, not necessarily obtained from
string compactifications, leading
to $AdS_4$ vacua that can be of interest.

Obtaining gauged $N=2, D=4$ supergravity seems to be important for
string theory compactifications since it is an intermediate step
between the more realistic $N=1$ models and the mathematically controllable
theories. Thus in the last decade there has been much literature
on the subject. An incomplete list of examples consists
of \cite{ferrara,aharony,gaunt,cassani,louis} and it is
straightforward to impose and solve the maximal supersymmetry
constraints in each case. In some cases the vacua have been
already discussed or must exist from general string
theory/M-theory considerations.

For example, it was found that the coset compactifications studied
in \cite{cassani} do not lead to $N=2$ supersymmetric configurations. This can also
be seen from imposing the constraints in section
\ref{summary}. In contrast, the compactification on $K3 \times
T^2/\mathbb{Z}_2$ presented in \cite{ferrara} does exhibit $N=2$
solutions with non-trivial hypermultiplet gaugings. The authors of \cite{ferrara} explicitly found $N=2$ Minkowski vacua by satisfying the
same susy conditions as in section \ref{summary}. From our analysis, it trivially follows that
also the pp-wave and the $AdS_2 \times S^2$ backgrounds are maximally supersymmetric. To check this, one only needs to verify \eqref{constr on pp-wave and r-b}, and this is satisfied due
to a vanishing $c$-tensor and the abelian gauging in the hypermultiplet
sector.

A similar example is provided by the  (twisted) $K3 \times T^2$ compactification of the heterotic string,
recently analyzed in \cite{louis}. For abelian gaugings, one can verify that the three zero scalar curvature vacua are present in these models.

We now turn to discuss the remaining models in more detail.

\subsection{M-theory compactification on $SU(3)$ structure
manifolds}\label{su(3) compactification}

There is a very interesting model for $N=2, D=4$ supergravity with
non-abelian gauging of the vector multiplet sector and non-trivial
$c$-tensor, arising from compactifications of M-theory on
seven-manifolds with $SU(3)$ structure \cite{aharony} (more
precisely, they consider Calabi-Yau (CY) threefolds fibered over a
circle). For the precise M-theory set-up, we refer the reader to \cite{aharony}; here we only discuss the relevant data for analyzing the maximal supersymmetry
conditions:

\begin{itemize}
\item the vector multiplet space can be parametrized by special coordinates, $X^\Lambda = (1, t^i = b^i + i v^i)$ and prepotential
\begin{equation}
F(X) = - \frac{1}{6} \kappa_{i j k} \frac{X^i X^j X^k}{X^0}\ ,
\end{equation}

  with the well-known triple intersection numbers $\kappa_{ijk}$ that depend on the particular choice of the CY-manifold.
  This gives the K\"{a}hler potential
  \begin{equation}\label{aharony example kaehler}
    {\cal K} = - \log \Big[ \frac{i}{6} \kappa_{i j k} (t^i-\bar{t}^i) (t^j-\bar{t}^j) (t^k-\bar{t}^k)\Big] \equiv - \log \mathrm{Vol}\ ,
  \end{equation}
  where $\mathrm{Vol}$ denotes the volume of the compact manifold. The gauge group is non-abelian with  structure
  constants
  \begin{equation}
  f_{\Lambda\Sigma}{}^0 = 0 = f_{i j}{}^k, \qquad f_{i 0}{}^j = - M_i^j\ ,
  \end{equation}
  and a  $c$-tensor whose only non-vanishing components are
  \begin{equation}
  c_{i, j k} = \frac{1}{2} M_i^l \kappa_{l j k}\ .
  \end{equation}
The constant matrix $M_i^j$ specifies the Killing vectors and moment-maps of the special K\"ahler manifold:
  \begin{equation}\label{aharony killing vectors}
    k^j_0 = - M^j_k t^k\ , \qquad \qquad k^j_i = M_i^j\ ,
  \end{equation}
  and
    \begin{equation}\label{aharony moment maps}
      P_0 = - M_i^j t^i \partial_j {\cal K}\ , \qquad \qquad P_i = M_i^j \partial_j {\cal K}\ .
    \end{equation}
    Not for any choice of $M_i^j$ is the Killing equation satisfied. As explained in \cite{aharony}, this is only the case when the relation \eqref{cyclicity} holds. This also ensures that \eqref{struct const condition} is satisfied, as one can easily check.

\item generally in this class of compactifications there always appear hypermultiplet scalars, but there is no gauging of this sector,
so the Killing vectors and the moment maps $P^x_{\Lambda}$ are
vanishing.
\end{itemize}

The scalar potential in this case reduces to the simple formula
  \begin{equation}\label{aharony scalar potential}
    V = - \frac{8}{\mathrm{Vol}^2} M_i^k M_j^l \kappa_{k l m} v^i v^j v^m\ ,
  \end{equation}
  which is positive semi-definite.

Analyzing the susy conditions is rather straightforward. Since $P^x = 0$, the
only allowed $N=2$ vacua are the ones with zero-scalar curvature. What is left for us to check are the conditions
$k^i_{\Lambda} \bar{L}^{\Lambda} = 0$ and $P_{\Lambda} L^{\Lambda}
= 0$. The latter is very easy to check and holds as an identity at every point in the special K\"ahler manifold. Also, it is equivalent to the relation $k^i_{\Lambda} L^{\Lambda} = 0$ which is  satisfied
whenever there exists a prepotential \cite{preABCDF}.
The condition $k^i_{\Lambda} \bar{L}^{\Lambda} = 0$ eventually
leads to
\begin{equation}\label{aharony susy condition}
     \frac{M^i_j (t^j - \bar{t}^j)}{\mathrm{Vol}} = 2 i \frac{M^i_j v^j}{\mathrm{Vol}} = 0\ , \qquad \forall i\ .
    \end{equation}
The solution to the above equation that always exists is the
decompactification limit when $\mathrm{Vol} \rightarrow \infty$.
The other more interesting solutions depend on the explicit form
of the matrix $M$. In case $M_i^j$ is invertible there are no
further solutions to \eqref{aharony susy condition}. On the other
hand, when $M$ has zero eigenvalues we can have $N=2$ M-theory
vacua, given by (a linear combination of) the corresponding zero
eigenvectors of $M$. For the supergravity approximation to hold,
one might require that this solution leads to a non-vanishing (and
large) volume of the CY. Each eigenvector will correspond to a
flat direction of the scalar potential, and with $V=0$ along these
directions. The case where the full matrix $M$ is zero corresponds
to a completely flat potential, the one of a standard M-theory
compactification on $CY \times S^1$ without gauging.

Thus it is clear that $M_i^j$ is an important object for this type
of M-theory compactifications and we now give a few more details
on its geometrical meaning \cite{aharony}. In the above class of
M-theory compactifications we have a very specific fibration of
the Calabi-Yau manifold over the circle. It is chosen such that
only the second cohomology $H^{(1,1)} (CY_3)$ is twisted with
respect to the circle, while the third cohomology $H^3 (CY_3)$ is
unaffected. Thus the hypermultiplet sector remains ungauged as in
regular $CY_3 \times S^1$ compactification, while the vector
multiplets feel the twisting and are gauged. This twisting is
parametrized exactly by the matrix $M$, as it determines the
differential relations of the harmonic (on the $CY_3$) two-forms:
\begin{equation}\label{aharony twisting matrix}
  {\rm d} \omega_i = M^j_i \omega_j \wedge {\rm d}z\ ,
\end{equation}
where $z$ is the circle coordinate.

Let us now zoom in on the interesting case when we have nontrivial
zero eigenvectors of $M$, corresponding to non-vanishing volume of
the CY. For a vanishing volume, or a vanishing two-cycle, the effective supergravity description
might break down due
to additional massless modes appearing in string
theory\footnote{For a detailed analysis of the possibilities in a
completely analogous case in five dimensions see \cite{mohaupt}}.
Therefore the really consistent and relevant examples for $N=2$
vacua are only those when the matrix $M$ is non-invertible with
corresponding zero eigenvectors that give nonzero value for every
$v^i$.

To illustrate this better, we consider a particular example, given
in section 2.5 of \cite{aharony}, of a compactification where the
$CY_3$ is a $K3$-fibration. In this setting one can explicitly
construct an $M$-matrix, compatible with the intersection numbers
$\kappa_{i j k}$. Here one can find many explicit cases where all
of the above described scenarios happen. As a very simple and
suggestive example we consider the 5-scalar case with $\kappa_{1 2 3}
= -1, \kappa_{1 4 4} = \kappa_{1 5 5} = 2$, and twist-matrix
\begin{equation}\label{aharony M-matrix}
M =\begin{pmatrix} 0 & 0 & 0 & 0 & 0 \\
0 & 4 & 0 & -2 & -2 \\
0 & 0 & -4 & 2 & 2 \\
0 & 1 & -1 & 0 & 0 \\
0 & 1 & -1 & 0 & 0
\end{pmatrix}\ .
\end{equation}
The general solution of $M \cdot \vec{v} = 0$ is
\begin{equation}\label{aharony solution}
\vec{v} =\lambda \begin{pmatrix} 1 \\ 0 \\ 0 \\ 0 \\ 0
\end{pmatrix}\ + \mu \begin{pmatrix} 0 \\ 1 \\ 1 \\ 2 \\ 0
\end{pmatrix}\ + \nu \begin{pmatrix} 0 \\ 1 \\ 1 \\ 0 \\ 2
\end{pmatrix}\ ,
\end{equation}
and the resulting volume is
\begin{equation}
\mathrm{Vol} = 8 \lambda \left( 2
\mu^2 + 2 \nu^2 + (\mu - \nu)^2 \right)\ ,
\end{equation}
which is clearly
positive semi-definite. In the case when either $\mu$ or $\nu$
vanishes we have a singular manifold that is still a solution to
the maximal supersymmetry conditions. When all three coefficients
(that are essentially the remaining unstabilized moduli fields)
are non-zero, we have a completely proper solution both from
supergravity and string theory point of view, thus providing an
example of $SU(3)$ structure compactifications with zero-curvature
$N=2$ vacua. This example can be straightforwardly generalized to
a higher number of vector multiplets, as well as to the lower
number of 4 scalars (there cannot be less than 4 vector multiplets
in this particular case).

It is interesting to note in passing that a special case of the
general setup described above was already known for more than
twenty years in \cite{Lauwers} (3.21), where $M_1^1 = -2, M_2^2 =
1,$ and $\kappa_{1 2 2} = 2$. It was derived purely from 4d
supergravity considerations, but it now seems that one can embed
it in string theory.

\subsection{Reduction of M-theory on Sasaki-Einstein$_7$}\label{SE_7}

There has been much advance in the last years in understanding
Sasaki-Einstein manifolds and their relevance for M-theory
compactifications, both from mathematical and physical
perspective. These spaces are good candidates for examples of the
$AdS_4/CFT_3$ correspondence and an explicit reduction to $D=4$
has been recently obtained in \cite{gaunt}. Originally the
effective lagrangian includes magnetic gauging and a scalar-tensor
multiplet, but after a symplectic rotation it can be formulated in
the standard $N=2$ formalism discussed here. After the dualization
of the original tensor to a scalar we have the following data for
the multiplets, needed for finding maximally supersymmetric vacua:
\begin{itemize}
  \item there is one vector multiplet, given by $X^{\Lambda} = (1, \tau^2)$ and $F(X) =
  \sqrt{X^0 (X^1)^3}$, leading to $F_{\Lambda} = (\frac{1}{2} \tau^3,\frac{3}{2} \tau^2)$ and K\"{a}hler potential
  \begin{equation}\label{SE_7 tensor kaehler}
    {\cal K} = - \log \frac{i}{2} (\tau - \bar{\tau})^3\ .
  \end{equation}
  There is no gauging in this sector, i.e. $k^i_{\Lambda} = 0$ and $P_{\Lambda} = 0$ for all $i,
  \Lambda$. This also means that both $f_{\Lambda \Sigma}{}^{\Pi}$ and $c_{\Lambda, \Sigma
  \Pi}$ vanish.
  \item the hypermultiplet scalars are $\{ \rho, \sigma, \xi, \bar{\xi} \}$ ($\rho$ and $\sigma$ are real, and $\xi$ is complex) with the universal hypermultiplet metric:
  \begin{equation}\label{quaternion metric SE_7 tensors}
    {\rm d}s^2 = \frac{1}{4 \rho^2} {\rm d} \rho^2 + \frac{1}{4
    \rho^2}\left({\rm d} \sigma - i(\xi {\rm d} \bar{\xi}-\bar{\xi} {\rm d} \xi ) \right)^2 + \frac{1}{\rho} {\rm d} \xi
    {\rm d} \bar{\xi}\ .
  \end{equation}

  We have an abelian gauging, given by (as there are no Killing vectors in the vector multiplet sector, we drop the tilde on the Killing vector fields in the hypermultiplet sector):
  \begin{equation}\label{killing vector SE_7 tensors}
    \tilde{k}_0 = 24 \partial_{\sigma} -4 i (\xi \partial_{\xi} - \bar{\xi}
    \partial_{\bar{\xi}})\ ,\qquad \tilde{k}_1 = 24 \partial_{\sigma}\ ,
  \end{equation}
and the moment maps, calculated in \cite{gaunt}, are
  \begin{equation}
   \nonumber P_0^1 = -4 \rho^{-1/2} (\xi + \bar{\xi})\ , \quad P_0^2 = 4 i \rho^{-1/2} (\xi - \bar{\xi})\ , \quad P_0^3 = -\frac{12}{\rho} + 4 \left(1 - \frac{\xi
   \bar{\xi}}{\rho}\right)\ ,
   \end{equation}
   \begin{equation}\label{moment map SE_7 tensors}
   P_1^1 = 0, \qquad P_1^2 = 0\ , \qquad P_1^3 = -\frac{12}{\rho}\ .
  \end{equation}
\end{itemize}
We can now proceed to solving the maximal supersymmetry
constraints. The conditions involving vector multiplet gauging are
satisfied trivially, while from $\tilde{k}^u_{\Lambda} L^{\Lambda}
= 0$ we obtain the conditions $\xi = \bar{\xi} = 0$ and $1 +
\tau^2 = 0$. Therefore $\tau = i$ (the solution $\tau = -i$ makes
the K\"{a}hler potential ill-defined) and ${\cal K} = - \log 4$.
However, not all the moment maps at this vacuum can be zero
simultaneously, leaving $AdS_4$ as the only possibility for a
$N=2$ vacuum solution. One can then see that $\epsilon_{x y z} P^y
\overline{P^z} = 0$ is satisfied, so the only remaining condition
is $P^3_{\Lambda} f^{\Lambda}_{\tau} = 0$. This fixes $\rho = 4$.
Therefore we have stabilized all (ungauged) directions in moduli
space: $\xi = \bar{\xi} = 0, \tau = i, \rho = 4$. The potential is
nonzero in this vacuum since $P^3 = 2$, which means the only
possibility for the space-time is to be $AdS_4$ with vanishing
field strengths. This is indeed expected since $SE_7$
compactifications of M-theory lead to an $N=2$ $AdS_4$ vacuum, the
one just described by us in the dimensionally reduced theory.

One can verify that this vacuum is stable under deformations in
the hypermultiplet sector of the type discussed in
\cite{antoniadis,lilia}. To show this, first observe that the
condition $\tilde{k}^u_{\Lambda} L^{\Lambda} = 0$ for $u = \xi$
always ensures vanishing $\xi$. Secondly, one may verify that the
deformations to the quaternionic moment maps are proportional to
$\xi$, and hence the remaining $N = 2$ conditions from section
\ref{2.4.1} are satisfied. It would be interesting to understand
if this deformation corresponds to a perturbative one-loop
correction in this particular type of M-theory compactification.

\subsection{Other gaugings exhibiting $AdS_4$ vacua}

Another example of an $AdS_4$ supersymmetric vacuum can be
obtained from the universal hypermultiplet. In the same
coordinates $\{\rho,\xi,\bar{\xi},\sigma\}$ as used in the
previous example, the metric is again given by \eqref{quaternion
metric SE_7 tensors}. This space has a rotational isometry acting
on $\xi$ and $\bar{\xi}$, given by ${\tilde k}_1-{\tilde k}_0$ in
the notation of \eqref{killing vector SE_7 tensors}. We leave the
vector multiplet sector unspecified for the moment, and gauge the
rotation isometry by a linear combination of the gauge fields
$A_\mu^\Lambda$. This can be done by writing the Killing vector as
\begin{align*}
\tilde k^u_\Lambda = \alpha_\Lambda (0,i \xi,-i \bar{\xi},0)\ ,
\end{align*}
for some real constant parameters $\alpha_\Lambda$. The
quaternionic moment maps can be computed to be
\begin{align*}
  P^x_\Lambda = \alpha_\Lambda \left(\frac{\xi + \bar{\xi}}{\sqrt
  \rho},\frac{i (\xi - \bar{\xi})}{\sqrt{\rho}},1 - \frac{\xi
  \bar{\xi}}{\rho}
  \right)\ .
\end{align*}
It can be seen that there are no points for which
${P}^x_\Lambda=0, \forall x$, so this means that only $AdS_4$
$N=2$ vacua are possible. To complete the example, we have to
specify the vector multiplet space, and solve the conditions
${P}^x_\Lambda f^\Lambda_i=0$ and $\tilde k^u_\Lambda L^\Lambda
=0$. The latter can be solved as $\xi = \bar{\xi} = 0$, and then
also $\epsilon^{xyz}P^y\overline{P^z}=0$. The first one then
reduces to $\alpha_\Lambda f^\Lambda_i = 0$. This condition is
trivially satisfied when e.g. $n_V=0$. A more complicated example
is to take the special K\"ahler space of the previous subsection
with no gauging in the vector multiplet sector. There is one
complex scalar $\tau$, a section $X^\Lambda = (1,\tau^2)$ and a
prepotential $F = \sqrt{X^0 (X^1)^3}$. We then find a solution for
$\tau = i \sqrt{\frac{- 3 \alpha_0}{\alpha_1}}$, under the
condition that $\alpha_0$ and $\alpha_1$ are non-vanishing real
constants of opposite sign. More complicated examples with more
vector multiplets may be constructed as well. It would be
interesting to study if such examples can be embedded into string
theory.

A similar situation arises in the absense of hypermultiplets. As
mentioned in the end of section \ref{hyperinos}, we can have
nonvanishing moment maps that can be chosen as $P^x_{\Lambda} =
\alpha_{\Lambda} \delta^{x 3}$. Then we again need to satisfy the
same condition $\alpha_\Lambda f^\Lambda_i = 0$ as above, and we
already discussed the possible solutions.

\section*{Acknowledgments}
We thank R. D'Auria, B. de Wit, S. Katmadas, A. Micu, and T.
Ort\'{i}n for useful discussion and correspondence, and E.
Plauschinn for his comments on the script. S.V. thanks the Simons
Center for Geometry and Physics at Stony Brook for hospitality.
The work of K.H. is supported by the Huygens Scholarship Programme
of NUFFIC.

\appendix

\section{Notation and conventions}\label{AppA}
We mainly follow the notation and conventions from~\cite{ABCDF}.
The action is defined by $S = \int \sqrt{|g|} \mathcal L$. We
start with the (ungauged) Lagrangian, whose Einstein-Hilbert and
scalar derivative terms read
\begin{align}\label{lagrangian}
\mathcal L = \frac 1 2 R + g_{i\bar \jmath} \partial_{\mu} z^i
\partial^{\mu} z^{\bar \jmath} + h_{uv} \partial_{\mu} q^u \partial^{\mu} q^v\ .
\end{align}

We set the Newton constant $\kappa^2=1$. As in~\cite{ABCDF}, we
use a $\{+,-,-,-\}$ metric signature. To get positive kinetic
terms for the scalars, we have to choose $g_{i\bar \jmath}$ and
$h_{uv}$ positive definite.

We compute the Riemann curvature as follows\footnote{Note
that this definition, when applied to the Riemann curvature of the
quaternionic manifold, differs with a
factor of $2$ compared with~\cite{ABCDF,D'Auria}. As a consequence,
there one has $R(h_{uv}) = - 4 n (n+2)$.}
\begin{align*}
{R^\rho}_{\sigma\mu\nu} &= \epsilon \left[ \partial_\mu\Gamma^\rho_{\nu\sigma}
    - \partial_\nu\Gamma^\rho_{\mu\sigma}
    + \Gamma^\rho_{\mu\lambda}\Gamma^\lambda_{\nu\sigma}
    - \Gamma^\rho_{\nu\lambda}\Gamma^\lambda_{\mu\sigma} \right]\ ,\\
R_{\mu\nu}&={R^\rho}_{\mu\rho\nu}\ , \quad R=g^{\mu\nu}R_{\mu\nu}\
,
\end{align*}
where $\epsilon = 1$ for Riemann spaces (the quaternionic and
special K\"ahler target spaces) and $\epsilon = -1$ for Lorentzian
spaces (space-time). The overall minus sign in the latter case is
needed to give $AdS$ spaces a negative scalar curvature. This
gives a sphere in Euclidean space (with signature $\{+,+,+,+\}$) a
positive scalar curvature.

The spin connection
enters in the covariant derivative
\begin{align*}
  D_\mu &= \partial_\mu - \frac 14 \omega_\mu^{ab} \gamma_{ab}\ ,\\
  \omega_\mu^{ab}  &= \frac 12 e_{\mu c}\left(\Omega^{cab} -
    \Omega^{abc} - \Omega^{bca} \right)\ ,\\
\Omega^{cab} &= \left(e^{\mu a} e^{\nu b} - e^{\mu b} e^{\nu
    a}\right) \partial_\mu e^c{}_\nu\ .
\end{align*}
The Lagrangian \eqref{lagrangian} is only supersymmetric if the
Riemann curvature of the hypermultiplet moduli space satisfies $
R(h_{uv}) = -8 n (n+2)\ , $ where $n$ is the number of
hypermultiplets, so the dimension of the quaternionic manifold is $4n$
(in applications to the universal hypermultiplet, we have $n=1$
and hence $R= - 24$).

Our conventions for the sigma matrices follow \cite{ABCDF}; in
particular they are symmetric and satisfy
$\left(\sigma^{xAB}\right)^* = -{\sigma^x}_{AB}$, and we have the
relation
\begin{align*}
  \sigma^x_{AB} \sigma^{yBC} = -{\delta_A^C} \delta^{xy} + i \epsilon_{AD}
  \epsilon^{xyz} \sigma^{zDC}\ .
\end{align*}
Indices are raised and lowered, on bosonic quantities, as
\begin{equation}
\epsilon_{AB}V^B=V_A\ ,\qquad \epsilon^{AB}V_B=-V^A\ .
\end{equation}
As mentioned in the main text, all fermions with upper $SU(2)_R$
index have negative chirality and all fermions with lower index
have positive chirality. We set $\gamma_5$ to be purely imaginary
and then complex conjugation interchanges chirality.

\section{Moment maps and Killing vectors on special K\"ahler manifolds}\label{AppB}

In this appendix, we present some further relevant formulae that
are used in the main body of the paper. First, we have defined the
moment maps on the special K\"ahler manifold as follows. Given an
isometry, with a symplectic embedding \eqref{Gtransf-sections}, we
can define the functions
\begin{equation}\label{mom-maps}
P_\Lambda\equiv i(k^i_\Lambda \partial_i{\cal K}+r_\Lambda)\ .
\end{equation}
Since the K\"ahler potential satisfies \eqref{Gtransf-K}, it is
easy to show that $P_\Lambda$ is real. From this definition, it is
easy to verify that
\begin{equation}
k^i_\Lambda=-ig^{i\bar \jmath}\partial_{\bar \jmath}P_\Lambda\ .
\end{equation}
Hence the $P_\Lambda$ can be called moment maps, but they are {\it
not} subject to arbitrary additive constants. Using \eqref{equiv2}
and \eqref{mom-maps}, it is now easy to prove the relation
\begin{equation}\label{equivar}
k^i_\Lambda g_{i\bar\jmath}k^{\bar\jmath}_\Sigma-k^i_\Sigma
g_{i\bar\jmath}k^{\bar\jmath}_\Lambda=if_{\Lambda\Sigma}{}^\Pi
P_\Pi\ ,
\end{equation}
also called the equivariance condition.

We can obtain formulas for the moment maps in terms of the
holomorphic sections. For this, one needs the identities
\begin{equation}\label{kdX}
k^i_\Lambda\partial_iX^\Sigma=-f_{\Lambda\Pi}{}^\Sigma
X^\Pi+r_\Lambda X^\Sigma\ ,\qquad
k_\Lambda^i\partial_iF_\Sigma=c_{\Lambda,\Sigma\Pi}X^\Pi+f_{\Lambda\Sigma}{}^\Pi
F_\Pi+r_\Lambda F_\Sigma\ ,
\end{equation}
which follow from the gauge transformations of the sections, see
\eqref{Gtransf-sections}. Using the chain rule in
\eqref{mom-maps}, it is now easy to derive
\begin{equation}\label{mom-maps2}
P_\Lambda={\rm e}^{\cal K}\Big[f_{\Lambda\Pi}{}^\Sigma(X^\Pi{\bar
F}_\Sigma+F_\Sigma{\bar X}^\Pi)+c_{\Lambda,\Pi\Sigma}X^\Pi{\bar
X}^\Sigma\Big]\ ,
\end{equation}
and similarly
\begin{equation}\label{Kill-vect}
k^i_\Lambda =-ig^{i\bar \jmath}\Big[f_{\Lambda\Pi}{}^\Sigma({\bar
f}^\Pi_{\bar \jmath} M_\Sigma+{\bar
h}_{\Sigma\,\bar\jmath}L^\Pi)+c_{\Lambda,\Sigma\Pi}{\bar
f}^\Pi_{\bar\jmath} L^\Sigma\Big]\ ,
\end{equation}
where we introduced $M_\Lambda \equiv {\rm e}^{{\cal
K}/2}F_\Lambda$ and $h_{\Lambda\,i}\equiv {\rm e}^{{\cal
K}/2}(\partial_i+{\cal K}_i)F_\Lambda$. The Killing vectors
\eqref{Kill-vect} are not manifestly holomorphic. This needs not
be the case because otherwise we would have constructed isometries
for arbitrary special K\"ahler manifolds, since holomorphic vector
fields obtained from a (real) moment map solve the Killing
equation.

\textbf{Note added}: We now show
that $P_\Lambda L^\Lambda = 0$, following the discussion in the
appendix of the second paper in~\cite{klemm}.

We start from the consistency conditions on
the symplectic embedding of the gauge transformations,
equations \eqref{kdX}. We eliminate $r_\Lambda$ using
\eqref{mom-maps}, and rewrite them as
\begin{align}\label{deriv-transformations}
-f_{\Lambda \Pi}{}^\Sigma L^\Pi &= k^i_\Lambda f_i^\Sigma+i
P_\Lambda L^\Sigma\ ,\\
f_{\Lambda \Gamma}{}^\Sigma M_\Sigma + c_{\Lambda,\Gamma\Sigma}
L^\Sigma &= k^i_\Lambda h_{i|\Gamma} + i P_\Lambda M_\Gamma\ ,
\end{align}
with $h_{i|\Gamma} = {\rm e}^{K/2} D_i F_\Gamma$. Multiplication of the first equation with $M_\Sigma$ and the second
with $L^\Gamma$ and subtracting leads to
\begin{align}\label{eq:comp}
2   f_{\Lambda\Gamma}{}^\Sigma L^\Gamma M_\Sigma + c_{\Lambda, \Gamma\Sigma}
 L^\Gamma L^\Sigma = 0\ ,
\end{align}
where we have used the identity $f_i^\Sigma M_\Sigma - h_{i|\Gamma}
  L^\Gamma = 0$. Contracting equation~\eqref{mom-maps2} with $L^\Lambda$ and
using~\eqref{eq:comp} and~\eqref{cyclicity} one finds
\begin{align}
  P_\Lambda L^\Lambda = 0\ ,
\end{align}
as announced below equation~\eqref{PL=0} and~\eqref{constr on pp-wave
  and r-b}. Contrating the first equation of~\eqref{deriv-transformations} with $L^\Lambda$ gives
$  L^\Lambda k^i_\Lambda f_i^\Sigma = 0$. It follows from contracting with $\text{Im}\, \mathcal N_{\Gamma \Sigma}
f^\Sigma_{\bar \jmath}$ that
\begin{align}
  L^\Lambda k^i_\Lambda = 0\ .
\end{align}
Here we have used the special geometry identities on the period
matrix, see e.g. \cite{ABCDF}
\begin{equation}
f_i^\Lambda ({\rm Im}\,{\cal
N})_{\Lambda\Sigma} {\bar
f}_{\bar\jmath}^\Sigma=-\frac{1}{2}\,g_{i\bar\jmath}\ .
\end{equation}

\section{Commutators of supersymmetry tranformations}\label{AppC}
Equating \eqref{susy-gravi} to zero gives an expression for the
supercovariant derivative $\nabla_\mu \epsilon_A$ in terms of the
matrices $T^-_{\mu \nu}$ and $S_{AB}$. Applying this operator
twice gives
\begin{align*}
  \nabla_\nu \nabla_\mu \varepsilon_A =& -\epsilon_{AB} D_\nu T_{\mu\rho}^-
  \gamma^\rho \epsilon^B && \epsilon_{AB}\\
&+ T^-_{\mu\rho} \gamma^\rho T^+_{\nu\sigma} \gamma^\sigma
\varepsilon_A&&\mathbf 1_A{}^B\\
&+ig \epsilon_{AB} T^-_{\mu \rho} \gamma^\rho \gamma_\nu
(S_{BC})^*
\varepsilon_C && {{\sigma^x}_A}^B\\
&-ig \epsilon^{BC} T^+_{\nu \rho} \gamma^\rho \gamma_\mu S_{AB}
\varepsilon_C && {{\sigma^x}_A}^B\\
&-g^2 S_{AB} (S_{BC})^* \gamma_\mu \gamma_\nu \varepsilon_C\ , &&
\mathbf 1_A{}^B+{{\sigma^x}_A}^B
\end{align*}
where we have indicated the $SU(2)$ structure on the right side.
In~\eqref{comm-eps}, the commutator does not contain a part
proportional to $\epsilon_{AB}$. This implies $D_\rho T_{\mu \nu}
= 0$. Calculation of the commutator now gives
\begin{align*}
  [\nabla_\nu, \nabla_\mu] \varepsilon_A =&+ T^-_{\mu\rho} \gamma^\rho
T^+_{\nu\sigma} \gamma^\sigma
\varepsilon_A - (\mu \leftrightarrow \nu)\\
&+\frac g 2 \left(T^-_{\nu \rho} \gamma^\rho \gamma_\mu
 \overline P^x + T^+_{\nu \rho} \gamma^\rho \gamma_\mu
 P^x\right)  {{\sigma^x}_A}^C
\varepsilon_C  - (\mu \leftrightarrow \nu)\\
&- \frac {g^2}2 \left(P^x \overline {P^x} \delta_A^C -
P^x \overline
  {P^y} \epsilon^{xyz} {{\sigma^z}_A}^C \right) \gamma_{\mu\nu}
\varepsilon_C\ .
\end{align*}
We equate this to~\eqref{comm-eps}, where we
use~\eqref{graviphoton-identity} and the condition~\eqref{defG}:
\begin{align*}
  [\nabla_\mu, \nabla_\nu] \epsilon_A &= -\frac 14 {R_{\mu\nu}}^{ab}
\gamma_{ab}
  \epsilon_A- ig
  F^\Lambda_{\mu \nu} P_\Lambda \epsilon_A + ig
  {{\sigma^x}_A}^B F^\Lambda_{\mu \nu} P^x_\Lambda \epsilon_B\\
&= -\frac 14 {R_{\mu\nu}}^{ab}\gamma_{ab}
  \epsilon_A - ig
  F^\Lambda_{\mu \nu} P_\Lambda \epsilon_A - g \left(\overline
  {P^x} T^-_{\mu\nu} - P^x T^+_{\mu\nu}\right) {{\sigma^x}_A}^B
  \epsilon_B\ .
\end{align*}
Some algebra now yields the necessary and sufficient conditions to
match the terms proportional to $\sigma^x{}_A{}^B$:
\begin{align*}
  T^-_{\mu \nu} \overline{P^x} &= 0\\
  \epsilon^{xyz} P^y \overline {P^z} &= 0\ ,
\end{align*}
which give the first conditions of section 2.3. The other
conditions are obtained by comparing the parts proportional to
$\mathbf 1_A{}^B$.

\section{Metrics and field strengths}\label{AppD}
\begin{itemize}
\item$AdS_2 \times S^2$\\ The line element, in local coordinates
$\{t,x,\theta,\phi\}$, is
\begin{align*}
  {\rm ds}^2 = q_0^2 \left( {\rm d}t^2 - \sin^2 (t) {\rm d}x^2 - {\rm
      d}\theta^2 - \sin^2 (\theta) {\rm d}\phi^2 \right)\ ,
\end{align*}
where $q_0$ is a real, overall constant which determines the size
of both $AdS_2$ and $S^2$. From~\eqref{spacetime-Riemann} we find the
only non--vanishing components
\begin{align*}
  T^+_{tx} &= \frac 12 q_0 \sin(t) {\rm e}^{i\alpha}\ ,\\
  T^+_{\theta\phi} &= -\frac i 2 q_0 \sin(\theta) {\rm e}^{i\alpha}\ .
\end{align*}

\item The pp-wave\\ The line element of a four--dimensional
Cahen-Wallach space~\cite{CahenWallach}, in local coordinates
$\{x^-,x^+,x^1,x^2\}$, is given by
\begin{align*}
  {\rm ds}^2 = -2 {\rm d}x^+ {\rm d}x^- - A_{ij} x^i x^j ({\rm d}x^-)^2
  - ({\rm d}x^i)^2\ ,
\end{align*}
where $A_{ij}$ is a symmetric matrix. Conformal flatness requires
$A_{11} = A_{22}$ and $A_{12} = 0$. We denote $A_{11} = -\mu^2$ as
$A_{11}$ should be negative. This space is known as the pp-wave. From~\eqref{spacetime-Riemann} we find the
only non--vanishing components
\begin{align*}
  T^+_{x^- x^1} &= \frac \mu 2 {\rm e}^{i\alpha}\ ,\\
T^+_{x^- x^2} &= -i \frac \mu 2 {\rm e}^{i\alpha}\ .
\end{align*}

\end{itemize}

\end{document}